%% file: PaperRSI.tex
\documentclass[%
 aip,
 amsmath,amssymb,
 reprint,%
]{revtex4-1}

\usepackage{graphicx}	
\usepackage{xcolor} 
\usepackage{lineno}
\usepackage[utf8]{inputenc}
\usepackage[acronym]{glossaries}
\usepackage[pdftex]{hyperref}
\usepackage{color,hyperref}
\usepackage{placeins}
\usepackage{orcidlink}

\DeclareUnicodeCharacter{2009}{\,} 
\DeclareUnicodeCharacter{0229}{\c{e}} 

\newcommand{\bham}{
\affiliation{Institute for Gravitational Wave Astronomy 
\& School of Physics and Astronomy, University of 
Birmingham, Birmingham, B15 2TT, United Kingdom}}

\newcommand{\nikhef}{
\affiliation{Dutch National Institute for Subatomic Physics, Nikhef, 1098 XG, Amsterdam, Netherlands}}

\newcommand{\vu}{
\affiliation{Vrije Universiteit Amsterdam, 1081 HV, Amsterdam, Netherlands}}

\newcommand{\Hanford}{
\affiliation{LIGO Hanford Observatory, Richland, WA 99352, United States of America}}

\newcommand{\Livingston}{
\affiliation{LIGO Livingston Observatory, Livingston, LA 70754, United States of America}} 

\newcommand{\Stanford}{
\affiliation{Stanford University, Stanford, CA 94305, United States of America}}

\newcommand{\Cardiff}{
\affiliation{Cardiff University, Cardiff, CF24 3AA, United Kingdom}}

\newcommand{\RAL}{
\affiliation{STFC Rutherford Appleton Laboratory, Chilton Didcot, OX11 OQX, United Kingdom}}

\newcommand{\Glasgow}{
\affiliation{Institute for Gravitational Research, University of Glasgow, Glasgow, G12 8QQ, United Kingdom}}

\newcommand{\CIT}{
\affiliation{LIGO, California Institute of Technology, Pasadena, CA 91125, United States of America}}

\makeglossaries
\loadglsentries[main]{glossary}

\makeatletter
\def\@email#1#2{%
 \endgroup
 \patchcmd{\titleblock@produce}
  {\frontmatter@RRAPformat}
  {\frontmatter@RRAPformat{\produce@RRAP{*#1\href{mailto:#2}{#2}}}\frontmatter@RRAPformat}
  {}{}
}%
\makeatother
\begin{document}

\preprint{AIP/123-QED}

\title{Reducing controls noise in gravitational wave detectors with interferometric local damping of suspended optics}

\author{J van Dongen \orcidlink{0000-0003-0964-2483}} 
 \nikhef \vu \email[The author to whom correspondence may be addressed: ]{jvdongen@nikhef.nl}

\author{L Prokhorov\orcidlink{0000-0002-0869-185X}}
\bham

\author{S J Cooper\orcidlink{0000-0001-8114-3596}}
\bham

\author{M A Barton\orcidlink{0000-0002-9948-306X}}
\Glasgow

\author{E Bonilla\orcidlink{0000-0002-6284-9769}}
\Stanford

\author{K L Dooley\orcidlink{0000-0002-1636-0233}}
\Cardiff

\author{J C Driggers \orcidlink{0000-0002-6134-7628}}
\Hanford

\author{A Effler\orcidlink{0000-0001-8242-3944}}
\Livingston

\author{N A Holland\orcidlink{0000-0003-1241-1264}}
\nikhef \vu

\author{A Huddart}
\RAL

\author {M Kasprzack\orcidlink{0000-0003-4618-5939}}
\CIT

\author{J S Kissel\orcidlink{0000-0002-1702-9577}}
\Hanford

\author{B Lantz\orcidlink{0000-0002-7404-4845}}
\Stanford

\author{A L Mitchell\orcidlink{0000-0003-2521-8973}}
\nikhef \vu

\author{J O'Dell\orcidlink{0000-0001-9238-255X}}
\RAL

\author{A Pele\orcidlink{0000-0002-1873-3769}}
\CIT

\author{C Robertson}
\RAL

\author{C M Mow-Lowry\orcidlink{0000-0002-0351-4555}}
\nikhef \vu

\date{\today}


\begin{abstract}
Control noise is a limiting factor in the low-frequency performance of the \acrshort{LIGO} gravitational-wave detectors. In this paper we model the effects of using new sensors called \acrshort{HOQI}s to control the suspension resonances.  We show if we were to use \acrshort{HOQI}s, instead of the standard shadow sensors, we can suppress resonance peaks up to tenfold more while simultaneously reducing the noise injected by the damping system. Through a cascade of effects this will reduce the resonant cross-coupling of the suspensions, allow for improved stability for feed-forward control, and result in improved sensitivity of the detectors in the $10-20$\,Hz band. This analysis shows that improved local sensors such as \acrshort{HOQI}s should be used in current and future detectors to improve low-frequency performance.
\end{abstract}
\maketitle

\glsresetall


\section{Introduction}
\label{sec:intro}

Gravitational waves were predicted by Einstein's Theory of General Relativity, and were first observed in 2015 \cite{GW150914}.  Since then multiple events  \cite{O21,O3} have been detected by the Advanced \gls{LIGO} \cite{AdvancedLigo} and Advanced Virgo \cite{AdvancedVirgo} interferometers. These interferometers precisely measure the \gls{DARM} changes of the long (3-4\,km) arm cavities. Passing gravitational waves induce differential strain in the perpendicular arms, allowing interferometric detection.

The first detection of a binary neutron star inspiral \cite{BNS,BNSMMS} demonstrated the importance of gravitational wave detectors for multi-messenger astronomy. Gravitational wave detectors provide sky localization from triangulation with multiple detectors. Inspiral detection can provide early alerts for electromagnetic and particle observatories. 

Improvements to sensitivity at $10-20$\,Hz enable earlier detections, greater signal-to-noise ratios, and further astrophysical reach into space \cite{A+}. All events observed so far have been inspirals, where the frequency increases until the objects collide and merge. However, the signals have much longer duration at lower frequencies, with the time-until-merger proportional to $f^{-8/3}$. Improvements to low frequency sensitivity are therefore especially important for earlier detections and increasing the time in the measurement band, which in turn improves sky localisation \cite{5hz}.

One of the largest noise sources in earth-bound gravitational-wave detectors is ground vibration, which is ten orders of magnitude larger than the signal \cite{GW150914} and moves the optics of the interferometer. The \gls{LIGO} observatory has seismic isolation systems consisting of cascaded passive \cite{QuadUpdate} and active isolation \cite{ISI1, ISI2} to suppress this movement noise, and facilitate gravitational wave detection. Passive isolation is achieved through the use of pendula, and mass-spring-systems. Active isolation employs a blend of relative displacement and inertial sensors for feedback control of the passive isolation systems. 

Despite the success of these isolation systems in reducing direct vibration coupling, Advanced \gls{LIGO}'s low-frequency sensitivity is limited at $10-20$\,Hz by `global control noise' from the interferometer's \gls{asc} and the auxiliary \acrlong{LSC} systems \cite{O33}. 

Global controls keep the optics of the interferometer correctly placed and oriented relative to each other. Local controls, on the other hand, minimize the transfer of ground motion to an individual optic. 

In this paper we analyse how improved local sensors and controls can improve performance in a manner that improves the performance and predictability for global controls. Suspension chains with better local damping produce a quieter, simpler, and more stable plant, thereby reducing noise that is associated with non-linear, bi-linear, and non-stationary couplings that cannot currently be suppressed in post-processing \cite{O3,MachineNoise}.

There is a large and growing community of instrumentation development for 3G observatories (for a subset of proposed sensors and measurement methods see references \cite{DOSEM,COLLETTE201572,intorev,smetana2022compact,Gerberding:15,Miller:12,https://doi.org/10.48550/arxiv.2111.14355} and inertial sensors \cite{https://doi.org/10.48550/arxiv.2204.04150,10.1785/0120090136,ZHAO2020106959,8336722,doi:10.1063/1.4881936,Kohlenbeck:2018eit,Cooper_2022,Mow_Lowry_2019,VerticalInertial,doi:10.1063/5.0047069,rohtua,alma9920725267002321,Chang:20,RossThesis,lfsvm,doi:10.1063/1.4862816,DING2022113398,0047069,Ubhi_2021,Korth_2016}  ). In this paper we present a novel analysis of the projected quantitative effect of these instruments on the performance of a multi-stage suspension. This is the most detailed analysis of this kind. It includes the most important cross-couplings and all known input noise sources based on measurements from LIGO, and produces output metrics that are relevant for global interferometer controls. While there are currently no models that correctly predict the detector sensitivity based on local suspension performance, we qualitatively elaborate on the connection between local sensors and improved detector sensitivity at low frequencies. In particular, we show how local resonances below 3\,Hz result in noise in the $10-20$\,Hz region. Our results support the statement that improved damping is one of the elements required for breaking the `low-frequency wall' \citep{5hz}

The MIMO suspension model used in this work is a modification of previous models that includes the design parameters for the new \gls{BBSS}, which is the first Advanced LIGO suspension to receive a major upgrade \cite{APDR}. Previous suspensions show excellent agreement between the MIMO model and measured transfer functions, with the exception of cross-couplings, which are not included in the stiffness matrix \cite{SimulationComparison}. We include the state-space suspension model, the control filters, and scripts for generating plots as supplemental material.

The evaluation of improved damping for the \gls{BBSS} is a case-study applicable to other triple suspensions, and with some modifications, to the quadruple suspensions used for the test mass optics. Installing better sensors at the \gls{BBSS} would be an effective technology demonstration. 

\section{Presented Opportunity: The {LIGO} A+ Big Beamsplitter Suspension}
\label{sec:Opportunity} %

\begin{figure}[ht]
   \includegraphics[width=\linewidth]{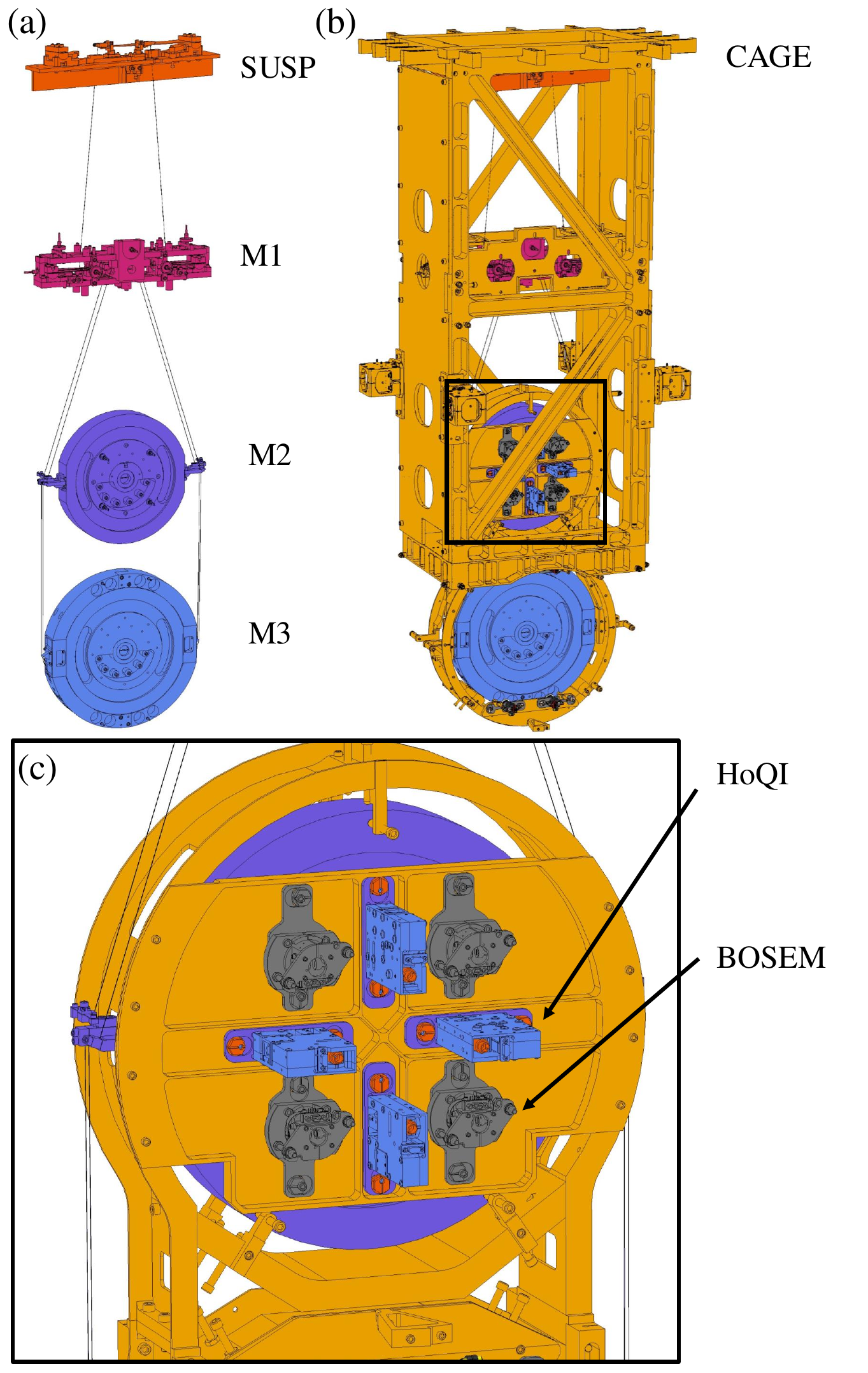}
   \caption{3D CAD rendering of the \gls{BBSS} triple. 
   \\ (a) the suspended masses; \acrfull{M1}, \acrfull{M2}, \acrfull{M3} that are mounted to the cage through the \gls{SUSP}. 
   \\ (b) the suspended masses together with the cage which is fixed to the \gls{ISI} platform. 
   \\ (c) detailed view showing the proposed sensor and actuator arrangement at \gls{M2}, showing \gls{BOSEM}s (used only for actuation) in the corners and \gls{HOQI}s (only capable of sensing) along the vertical and horizontal axes. }
    \label{fig:TripleSuspensions}
\end{figure}

\begin{figure}[ht]
   \includegraphics[width=\linewidth]{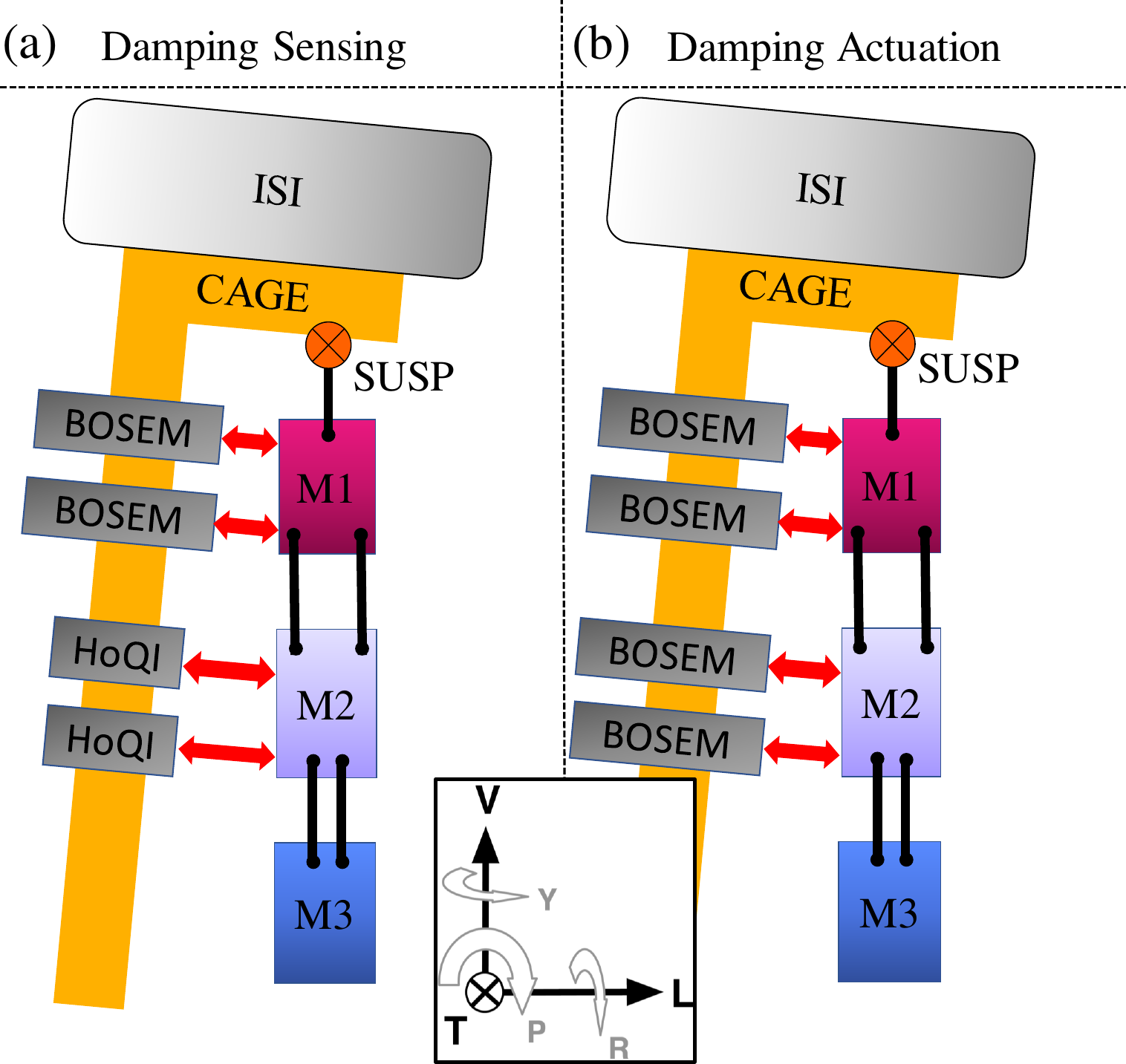}
   \caption{Simplified side-view schematic of the triple \gls{BBSS} showing the sensing and actuation points for damping the remainder of ground motion not suppressed by the \gls{ISI}.\\ (a) Overview of the sensors used for damping, at \gls{M1}, multiple \gls{BOSEM}s are mounted on the cage, measuring the relative displacement between the cage and \gls{M1} in all six degrees of freedom. Similarly, at \gls{M2} four \gls{HOQI}s are mounted on the cage, measuring the relative displacement between the cage and \gls{M2}, but only in \acrfull{L}, \acrfull{P}, and \acrfull{Y}.\\ (b) Overview of the actuators used for damping, at both \gls{M1} and \gls{M2} there are multiple \gls{BOSEM}s to actuate on the suspension. By default, the \gls{BOSEM}s at \gls{M1} are used for local sensing and control and provide static offsets for global control, and the \gls{BOSEM}s at \gls{M2} for actuation from global interferometer signals. We investigate using the \gls{BOSEM}s at \gls{M2} for actuation based on both local and global signals.}
    \label{fig:TripleSuspensionsSchematic}
\end{figure}

One of the \gls{LIGO} A+ upgrades, planned for after the upcoming fourth observing run, is the installation of a new, larger beamsplitter \cite{APDR}. This necessitates the new \acrfull{BBSS} that contains slots for optional \acrlong{HOQI}s (\acrshort{HOQI}s)\cite{HoQI} to be used as relative displacement sensors. Compact interferometric sensors \cite{intorev}, like \gls{HOQI}s  provide significant performance improvements \cite{HoQI} to the baseline \acrfull{BOSEM} \cite{Bosem,A+Bosem}. This makes the \gls{BBSS} a perfect test case for modeling how better sensors could affect the performance of the suspended optics. 

Figure \ref{fig:TripleSuspensions} shows an overview of the \gls{BBSS}, a cascade of 3 masses. The suspension system isolates the \gls{M3} from the residual ground motion of the \acrfull{ISI} platform. Every step of this chain reduces the motion transmitted to the lower mass, with an $f^{-2}$ power law above its pendulum frequency. The chain's resonances must be damped to reduce the \gls{RMS} motion of \gls{M3}. The damping system works by measuring and actuating between the rigid cage and the suspended stages, depicted in Figure \ref{fig:TripleSuspensionsSchematic}. The baseline LIGO damping design uses \gls{BOSEM}s \cite{Bosem} mounted on the cage to measure and actuate on the \acrfull{M1} (we call this `\gls{M1} \gls{BOSEM} damping'). 

\vfill\null

In this paper we present a detailed investigation of the damping and noise performance if \gls{HOQI}s are mounted between the \acrfull{M2} and the cage, while using the \gls{BOSEM}s for actuation. It is the first study showing that the improved sensitivity allows \gls{HOQI}s to be used for local control at a stage closer to the optic, resulting in more control authority and improved damping without disturbing the sensitivity in the critical measurement band above 10\,Hz. The combination of \gls{M2} \gls{HOQI} sensing and \gls{M2} \gls{BOSEM} actuation is referred to as `\gls{M2} \gls{HOQI} damping'. We will provide a detailed noise-budget breakdown of the contributions to optic motion for the \gls{BBSS} in both Length and Pitch for both the \gls{BOSEM} and \gls{HOQI} damping scenarios. Four targets were identified as crucial for control design: a stable controller that reduces the suspension resonance peaks, meeting 10\,Hz performance requirements and reducing \gls{RMS} motion.


\section{Simulating the {HOQI} damping performance}
\label{sec:Method} %

The purpose of the damping system is to lower the quality factor of the resonances of the suspension chain at its eigenfrequencies. This reduces the total motion at \gls{M3}, provides a simpler plant for implementing global interferometer controls, and reduces the \gls{RMS} motion of the suspended masses and corresponding non-linear effects. Active damping is preferred over passive damping to have more freedom in shaping the frequency response of the dissipation and to allow for fine-tuning after installation. In addition to damping performance, this system should not introduce noise into the sensing region of the detector (10\,Hz and above). To design the damping system we need models of the damping performance and an understanding of the noise contributions.

The following method is used to show the damping performance of \gls{M2} \gls{HOQI} damping in comparison to \gls{M1} \gls{BOSEM} damping. A Matlab model for simulating the damping performance of the \gls{BOSEM}s at \gls{M1} exists \cite{bbssmatlab, BBSSM}. This model provides an open-loop (undamped) state-space for the suspension dynamics, and includes feedback paths and damping filters for sensing and actuating at \gls{M1}. These produce the closed-loop \gls{BOSEM}-damped state-space.

We expanded this implementation to include feedback paths from \gls{HOQI}s sensors to \gls{BOSEM}s actuators, all at \gls{M2}, and designed appropriate new stable damping filters. This study is limited to the \gls{L} and \gls{P} degrees of freedom, which are strongly cross-coupled in the underlying mechanical equations of motion. Similar damping performance is expected for \gls{Y}.  

Three primary noise sources are identified: actuator noise, sensor noise, and inertial noise. 

For actuator noise we re-used the validated actuator noise model of the \gls{LIGO} beamsplitter suspension used in the current observing run \cite{BSFMactuator}, and updated the geometry and DAC to those planned for the \gls{BBSS}. For global control all actuators are used independent of the local controls, so in both the \gls{M2} \gls{HOQI} and \gls{M1} \gls{BOSEM} damped scenarios, actuator noise originates from both the \gls{M1} and \gls{M2} actuators. \gls{M3} actuator noise was not included since important design parameters are still missing.  Actuator noise is primarily caused by the DAC voltage noise, which generates force noise at the suspended masses. 

Sensor noise depends on the resolution of the individual sensors and their geometrical placement. Depending on the control filters and the loop gain, the sensor noise will be injected via actuators into the suspension. Both \gls{HOQI}s and \gls{BOSEM}s are relative position sensors. To measure translation and rotation, multiple sensors are used at different locations of \gls{M1} and \gls{M2}. Common sensor output corresponds to translational movement, and the noise is lower than that of an individual sensor due to the multiplicity of sensors. Differential sensor outputs correspond to rotation and the noise is higher than for an individual sensor due to the short lever arm. Scripts were used to obtain the sensing noise in each \gls{dof} based on the geometric placement of \gls{BOSEM}s for the \gls{M1} \gls{BOSEM} damping model  \cite{OSEMEUL} and modified for use with the \gls{M2} \gls{HOQI} damping model. 

Inertial noise consists of two parts. The first part is `\gls{ISI} noise', the transmission of motion from the \gls{ISI} (rigidly connected to \acrfull{SUSP}) to \gls{M3}. The second part is `cage noise'. 
At high frequencies, the masses \gls{M1}-\gls{M3} move much less than the cage, but the sensors can only measure the relative motions. To model this we projected the \gls{ISI} movement to the \gls{M1} and \gls{M2} sensor locations. 
It is non-trivial to accurately determine the inertial rotation of the \gls{ISI} at low frequencies, and we used recent predictions of expected tilt, synthesised from several sensors \cite{isiblend}.

Since the \gls{BBSS} hangs down from the \gls{ISI}, the motion of the cage due to rotation of the \gls{ISI} increases at each stage. However, damping performance increases substantially when measuring and actuating closer to the optic. With a total noise budget and the complete damped suspension model, the effect of \gls{M1} \gls{BOSEM} versus \gls{M2} \gls{HOQI} damping was compared quantitatively. The full matlab model of our simulations is shared online \cite{matlabscripts}.


\section{{HOQI} vs {BOSEM} damping performance}
\label{sec:Results} %

\begin{figure}[ht]
	   \includegraphics[width=\linewidth]{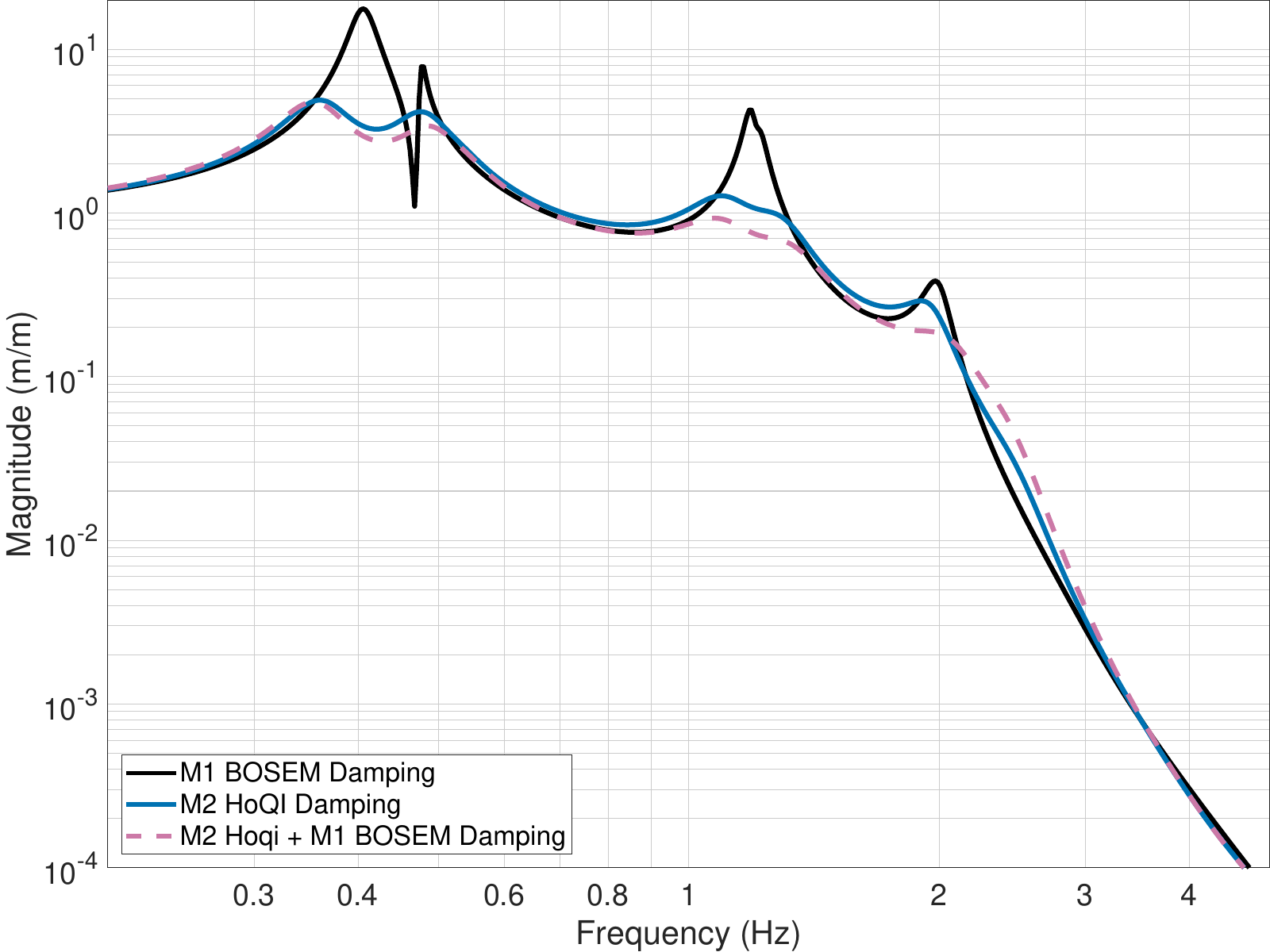}
	   \caption{In loop damped transfer function from \gls{ISI} translation to optic translation in the `\gls{L}' direction (parallel to the optical axis) for different input frequencies. A comparison is made between three local sensing and control configurations: only at \gls{M1} or \gls{M2}, and both together. Outside of the displayed range the responses were identical.}
	    \label{fig:LtoL}
	\end{figure}

\begin{figure}[!htb]
		   \includegraphics[width=\linewidth]{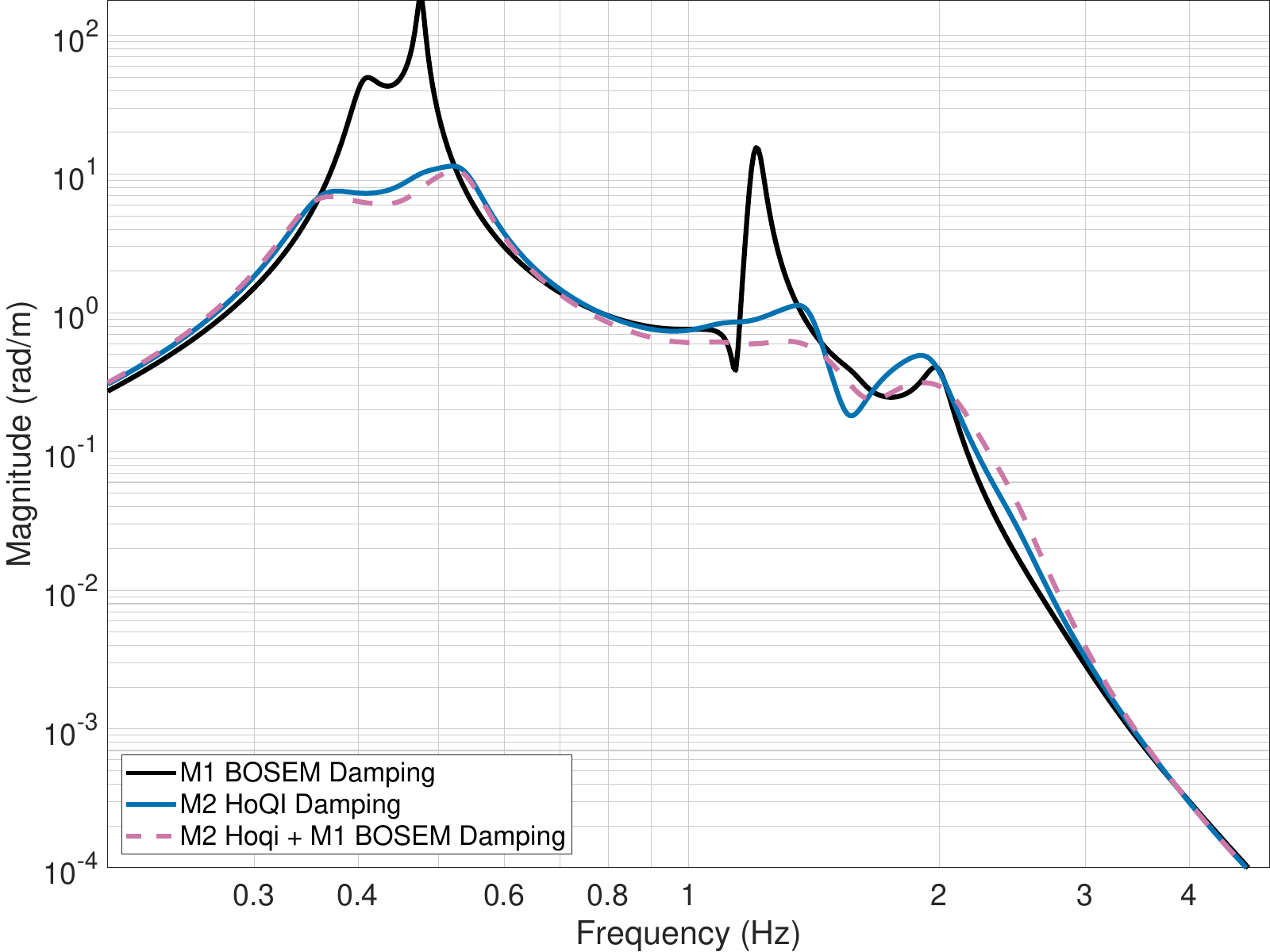}
		   \caption{In loop transfer function from \gls{ISI} translation to optic rotation in the `\gls{P}' direction. A comparison is made between three sensing and actuation configurations: only at \gls{M1} or \gls{M2}, and at both together. Outside of the displayed range the responses were identical.}
		    \label{fig:LtoP2}
	\end{figure}

\begin{figure}[!htb]
		   \includegraphics[width=\linewidth]{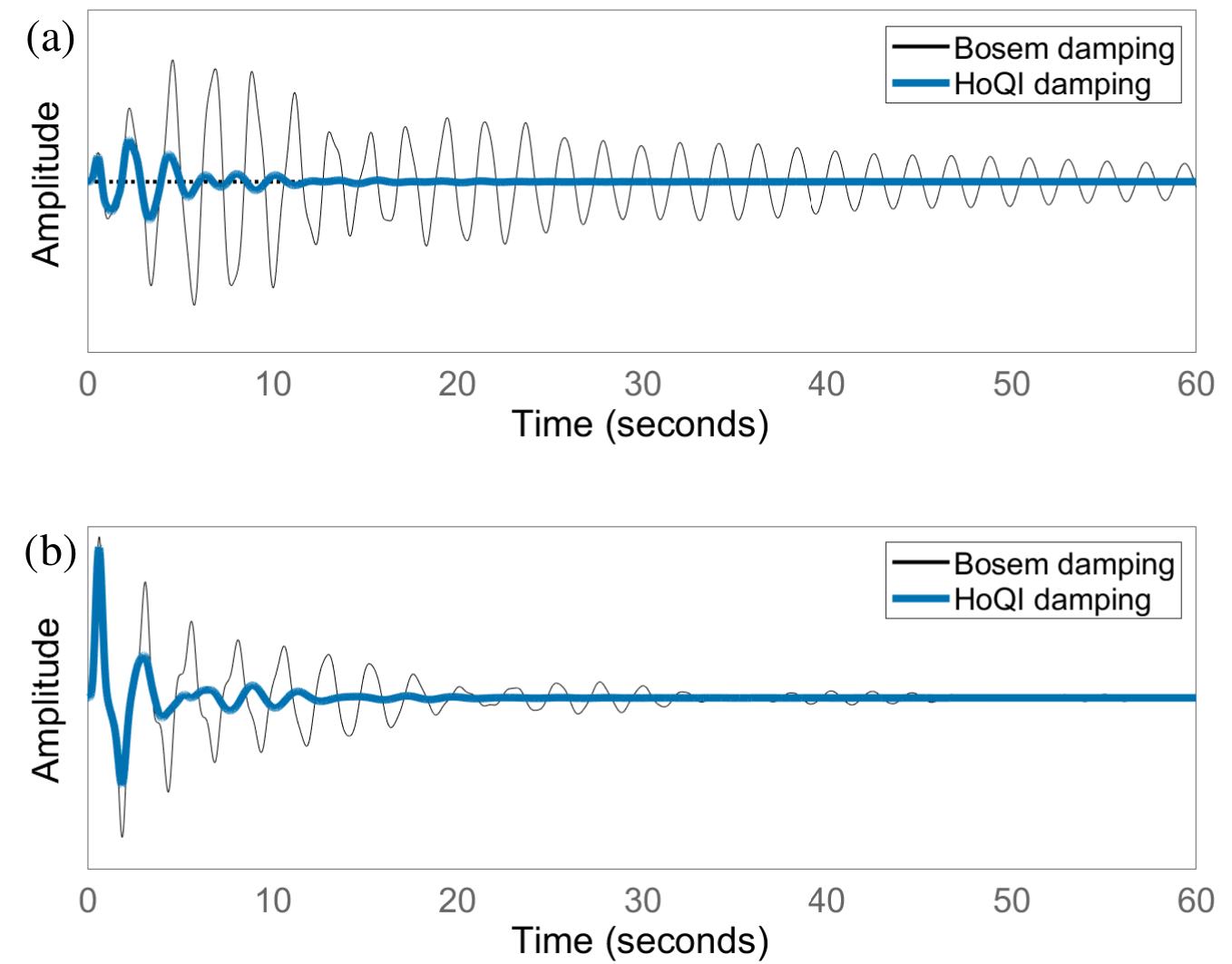}
		   \caption{Impulse response from \gls{ISI} `\gls{L}' (\gls{SUSP} drive location) resulting in (a) \gls{M3} `\gls{L}' movement and (b) \gls{M3} `\gls{P}' movement. Figures show purely the damped plant response, sensor and cage noise have not been taken into account. The \gls{M2} \gls{HOQI} damped system exhibits a much shorter transient response.}
		    \label{fig:ISIM3}
	\end{figure}

We close the control loops on the \gls{BBSS} model with sensing and actuation at \gls{M2}. Figure \ref{fig:LtoL} shows the transfer function of \gls{ISI} \acrlong{L} motion to \gls{M3} \acrlong{L} displacement with different damping configurations shown; the baseline sensing and damping at \gls{M1}, our proposed sensing and damping at \gls{M2}, and both damping methods simultaneously. It can be seen that the resonances around 0.4\,Hz, 1.1\,Hz and 1.8\,Hz are suppressed by up to a factor of eight with \gls{M2} \gls{HOQI} damping compared with (only) \gls{M1} \gls{BOSEM} damping. The results for \gls{M2} \gls{HOQI} damping, and \gls{M2} \gls{HOQI} + \gls{M1} \gls{BOSEM} damping are very similar in performance. The performance improvements in \acrfull{P} can be found in figure \ref{fig:LtoP2}. The inpulse responses from the \gls{ISI} to the mirror \gls{M3} in figure \ref{fig:ISIM3} show the improved settling time with the \gls{HOQI}-damped suspension compared to \gls{BOSEM}-damped. We have only considered configurations that are feasible for the \gls{BBSS}.
		
	\begin{figure}[ht]
	   \includegraphics[width=\linewidth]{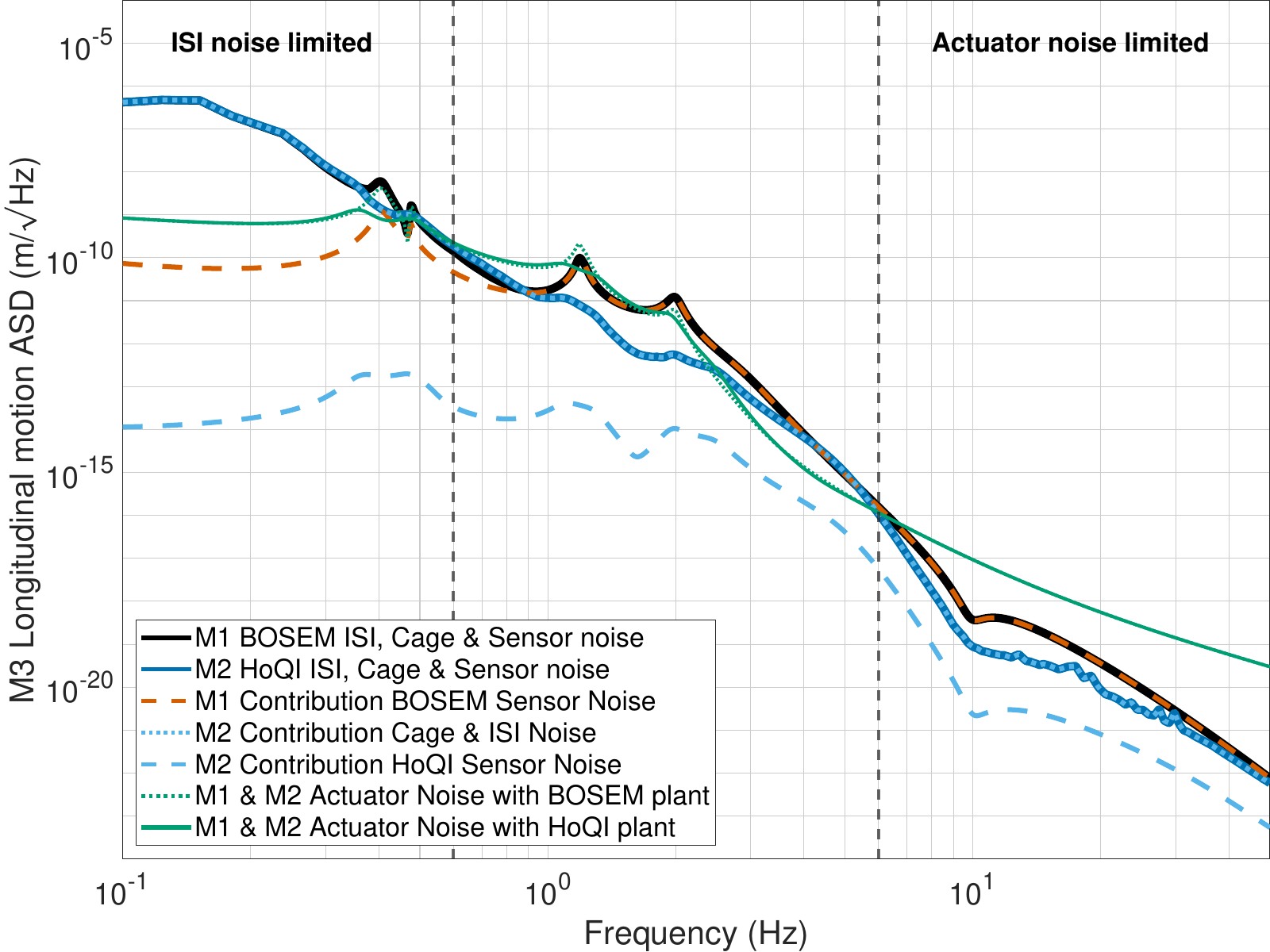}
	   \caption{Noise budget \gls{M3} translation in the `\gls{L}' direction,  displaying the Cage, ISI and Sensor noise when using \gls{BOSEM} damping at \gls{M1} (black and brown), or \gls{HOQI} damping at \gls{M2} (light and dark blue). And for both systems the actuation noise (green). Below 0.6\,Hz the noise budget is limited by \gls{ISI} movement, above 6 Hz it is limited by actuator noise}
	    \label{fig:NtoL}
	\end{figure}

	\begin{figure}[ht]
	   \includegraphics[width=\linewidth]{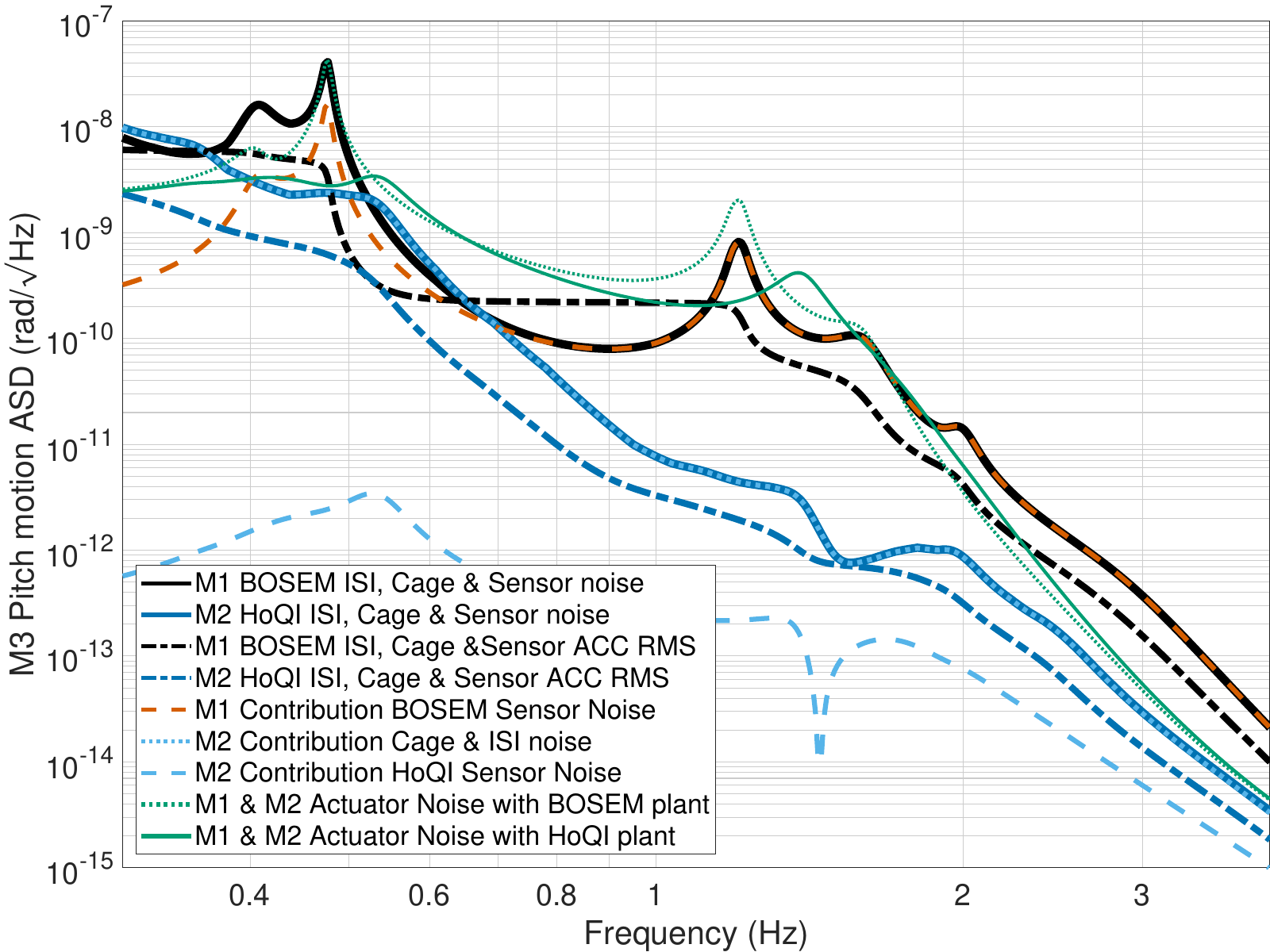}
	   \caption{Noise budget of \gls{M3} rotation in the `\gls{P}' direction, displaying the Cage, ISI and Sensor noise when using \gls{BOSEM} damping at \gls{M1} (black and brown), or \gls{HOQI} damping at \gls{M2} (light and dark blue). And for both systems the actuation noise (green). The figure is zoomed in around the resonance frequencies where the biggest differences occur.}
	    \label{fig:NtoP}
	\end{figure}

Figures \ref{fig:NtoL} and \ref{fig:NtoP} present the closed-loop noise budgets, showing different noise sources contributing to optic \acrfull{L} and \acrfull{P} movement with either \gls{M1} \gls{BOSEM} or \gls{M2} \gls{HOQI} damping. When excluding the actuator noise it can be seen that above 1\,Hz, the \gls{HOQI}-damped system always has a lower total noise than the \gls{BOSEM}-damped system. Above 1\,Hz the \gls{BOSEM}-damped system is limited by \gls{BOSEM} sensor noise, while the \gls{HOQI} damped system is always limited by \gls{ISI} and cage noise. Finally, the accumulated \gls{RMS} movement in the \acrfull{P} direction at 1\,Hz is a factor 60 lower for the \gls{M2} \gls{HOQI}-damped system when compared to the \gls{M1} \gls{BOSEM}-damped system. When taking into account actuator noise the noise budget of both methods is comparable.

The results show that using \gls{M2} \gls{HOQI}-damping obtains better damping performance without introducing more noise.  Figure \ref{fig:LtoL} shows that when using \gls{M2} \gls{HOQI}-damping, additional \gls{M1} \gls{BOSEM}-damping does not bring substantial improvements. \gls{M1} \gls{BOSEM}-damping has two limitations: sensor noise and dynamic coupling. For sensor noise, if the gain is increased, motion at the optic will increase rather than decrease due to the injection of BOSEM sensor noise. For dynamic coupling, only some fraction of the total kinetic energy couples to the top mass, creating an impedance-matching limit for damping. Increasing damping gain beyond this limit increases optic motion. Combining this information with the lower noise injection of the \gls{HOQI} system, as shown in figures \ref{fig:NtoL} and \ref{fig:NtoP} provides strong motivation for deactivating \gls{BOSEM}-damping in degrees of freedom where \gls{HOQI}-damping can be used: \acrfull{L}, \acrfull{P}, \acrfull{Y}. 

\gls{BOSEM} sensor noise and actuator noise are driving the noise budget above 1\,Hz in the \gls{M1} \gls{BOSEM}-damped system. Actuator noise, cage noise and inertial motion of the \gls{ISI} and the cage dominates the \gls{M2} \gls{HOQI}-damped budget, especially in \acrfull{L}. If the sensors were instead mounted on a suspended reaction chain, present in the quadruple suspensions  used for suspending the test masses of main arm-cavity optics \cite{AdvancedLigo}, the cage noise will be strongly attenuated. This will result in even better performance for the \gls{HOQI} damped system. Cage noise can be reduced with `sensor correction', a feed-forward technique designed to subtract the inertial motion of the \gls{ISI} from the \gls{HOQI} sensor output. This is possible using the inertial sensors on the \gls{ISI}, which measure motion in the relevant band with significant signal-to-noise ratio \cite{isiblend}. This analysis has also exposed that actuation noise is a dominant noise source around 1\,Hz and above 10\,Hz. To realise the full improvements possible with interferometric local sensors, actuation noise should be suppressed by more than two orders of magnitude at frequencies near 1.5\,Hz. 

It is difficult to further reduce the noise of LIGO's DACs and significant gains can only be made by reducing the actuation strength. In order to reduce the required actuation strength required for alignment, it is possible to offload static offsets in \acrfull{P} and \acrfull{Y} to stepper motors, as implemented successfully at Virgo \cite{payload}. Additionally, slow drift in the \acrfull{L} direction can be corrected further upstream from suspension chain, at either the \gls{ISI} or \gls{HEPI}. If the actuators only have to compensate for drifts on short timeframes, less movement and therefore less actuation force is needed resulting in less noise from the DAC propagating to the suspensions. 


\section{Impact of suspension damping on gravitational-wave detection}
\label{sec:Interpretation} %

Current simulation models don't correctly predict the \gls{LIGO} interferometers' total noise budget for the $10-20$\,Hz region. Steps are being taken to model non-linear couplings into \gls{DARM} \cite{Lightsaber} and to use machine learning for non-linear noise prediction and suppression \cite{MachineNoise,MachinesSensitivity}. 

\begin{figure}[ht]
   \includegraphics[width=\linewidth]{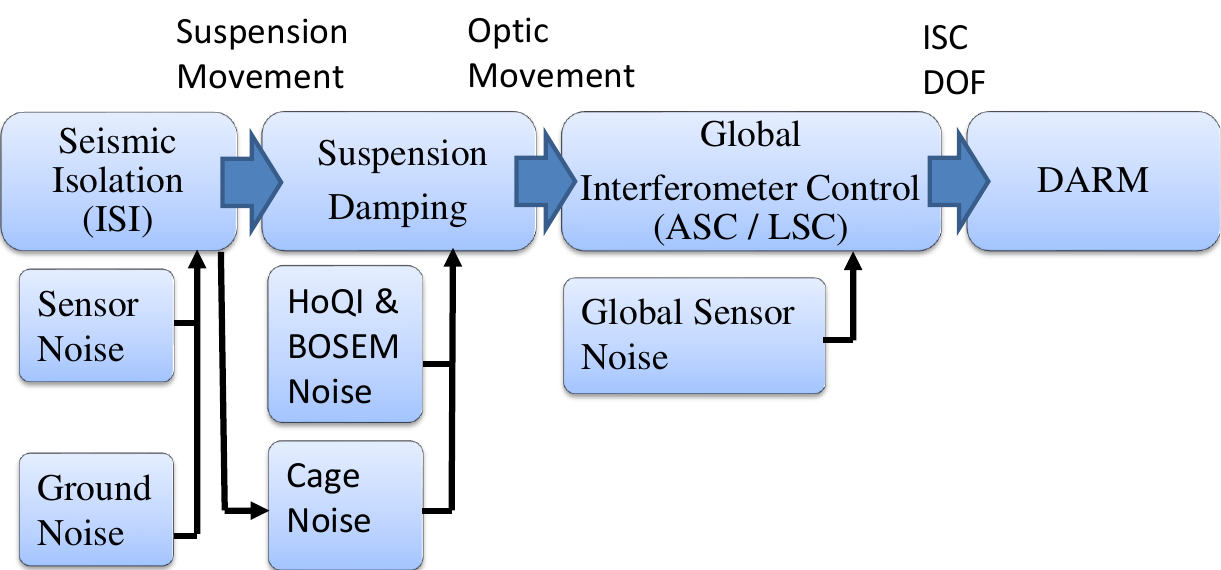}
   \caption{Simplified flow chart of the different stages of control and types of noise injections. The \gls{ISI} actively lowers the ground movement propagating to the suspensions. Measurement noises are seen as actual motion such that counteracting this perceived motion will cause actual motion. Sensor noise and ground noise result in \gls{ISI} noise. \gls{ISI} noise, cage noise and \gls{HOQI} and \gls{BOSEM} noise aggregate to create optic movement. Optic movement and global sensor noise (such as shot noise) cause movement between optics, which couples to \gls{DARM}. In the $10-20$\,Hz region these effects limit the detector performance.} 
    \label{fig:ControlNoise}
\end{figure}

Further efforts are being made to iterate on control filter design by closing an over-arching loop, with detailed technical noise projections \cite{closedloop}. Figure \ref{fig:ControlNoise} illustrates some of the interconnections between distinct control systems. An integrated loop over the different control systems would allow for quickly evaluating the \gls{DARM} performance gains seen by improving one element in the chain.

As these models are currently under development, we aren't able to quantify the improvements of better damping (for single or multiple optics) on \gls{DARM}. Instead we outline the mechanisms that link suspension dynamics and optic motion with \gls{DARM} sensitivity.

\begin{figure}[ht]
   \includegraphics[width=\linewidth]{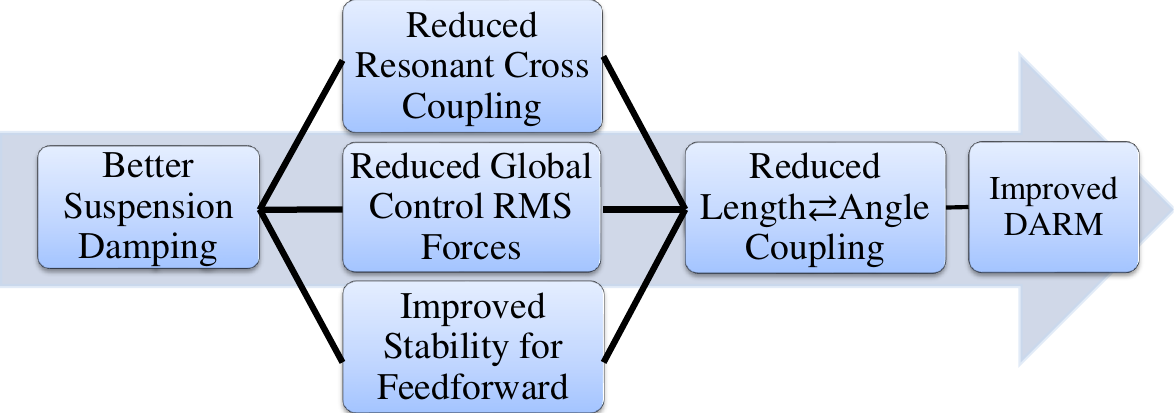}
   \caption{A simplified flow chart sketching how better suspension damping can improve \gls{DARM} sensitivity. Better suspension damping reduces the \gls{RMS} motion of the optic, and (especially) reduces eigenmode resonance peaks, making the plant easier to control. 
When global sensor noise in auxiliary degrees of freedom is limiting the performance, the reduced \gls{RMS} motion allows for auxiliary control bandwidths to be lowered with the same closed-loop motion. Lower control bandwidth reduces the injection of optical sensor noise. Lower resonance peaks make a more robust plant, allowing for more accurate feed-forward and more aggressive roll-off. Finally, energy in resonance peaks is strongly cross-coupled between eigenmodes that are close in frequency. Damping therefore reduces the strength of the mechanical cross-coupling between resonances. Combined, these improvements result in improved \gls{L}-to-angle and angle-to-\gls{L} coupling, which will result in better gravitational wave sensitivity.}
    \label{fig:Darm}
\end{figure}

Reducing the \gls{RMS} motion of \gls{LIGO}'s suspended optics allows for a reduction of interferometric sensing and control bandwidths, and therefore reduced control forces and lower optical sensor noise injection. Lower resonance peaks mean lower cross-coupling into different degrees of freedom. A better damped \gls{BBSS} (evidenced by `smoother' transfer functions), enables more robust and simpler global control options, and less of a need for frequency dependent features, which affect phase. Better damping (figure \ref{fig:ISIM3}) and `smoother' phase features should improve the quality and stability of the feed-forward, used to de-couple degrees of freedom.

The effects of having multiple better damped and therefore easier to control core optics are outlined conceptually in figure \ref{fig:Darm}. From interferometric \gls{asc} studies \cite{ASC} it is known that all interferometer degrees of freedom impact \gls{DARM}. Improvements in any part of this chain can lower the `technical noise' contributions to \gls{DARM}. Figure \ref{fig:Darm} stems from strategies which have helped resolve problems in the $20-40$\,Hz range \cite{SAMSRCL2}, and investigations into a main offender for the $10-20$\,Hz region \cite{SRCLMysteries}. 

The model adaptations we have made can be readily transferred to other triple suspensions in future upgrades. Initial studies showing the benefits for quadruple suspensions have been performed \cite{QuadDamp} and we expect that similar results can be obtained for all core suspended optics. The increased sensitivity of \gls{HOQI}s deeply suppresses the contribution of sensor noise and the inertial movement of the cage (cage noise) is expected to dominate the motion of the optic for \gls{BBSS} triple suspension. As such, lower total noise is anticipated if \gls{HOQI}s are placed on the suspended `reaction chain' of the quadruple suspensions that support the test masses. Even without a precise quantification, which is necessarily dependent on the (evolving) status of the interferometer, we have linked previously successful strategies that provide causal evidence that reduced motion via better damping should result in reduced noise in \gls{DARM} at frequencies between 10 and 30\,Hz.

\section{Conclusions}
\label{sec:conclusions} 

The \acrlong{BBSS} design has slots for installing \gls{HOQI}s, providing the potential to test improved sensors (and as such better damping) for this suspension, and act as a technology demonstrator for future upgrades. We have simulated the damping performance if \gls{HOQI}s are fitted and used the resulting performance as a case study for improved sensors in all suspensions. This kind of analysis is crucial for determining the improvements that can be realised in the detector, including existing limits imposed by other systems.

We have shown that by using \gls{HOQI} damping at \gls{M2}, instead of \gls{BOSEM} damping at \gls{M1}, we can reduce the resonance peaks of the plant by a factor of up to eight while simultaneously reducing the motion of the suspended optic. Reduced \gls{RMS} motion means lower global control bandwidth, reduced control forces, and less non-linear, bi-linear and non-stationary couplings. 
A simpler plant allows for more flexible and robust global control options. This better plant stability permits improvements to the feed-forward control. These aspects will improve the sensitivity of the \gls{LIGO} interferometers in the control-noise-limited $10-20$\,Hz region.



\begin{acknowledgments}
\label{sec:Acknowledgement}
We thank Norna Robertson and the Advanced \gls{LIGO} Suspensions team for work developing the \gls{BBSS} dynamics model. The authors gratefully acknowledge the support of the United States National Science Foundation (NSF) for the construction and operation of the LIGO Laboratory and Advanced LIGO. LIGO was constructed by the California Institute of Technology and Massachusetts Institute of Technology with funding from the United States NSF, and operates under cooperative agreement PHY-1764464, Advanced LIGO was built under award PHY-0823459. The authors acknowledge the support of the Institute for Gravitational Wave Astronomy at the University of Birmingham, STFC grants `Astrophysics at the University of Birmingham’ grant ST/S000305/1 and `The A+ upgrade: Expanding the Advanced \gls{LIGO} Horizon` ST/S00243X/1. The support for Cardiff University grants; Leverhulme Trust: PLP-2018-066 and UKRI | Science and Technology Facilities Council (STFC): ST/V005618/1.  This project has received funding from the European Research Council (ERC) under the European Union's Horizon 2020 research and innovation programme (grant agreement No. 865816).
\end{acknowledgments}

\section*{Data Availability Statement}

The data that support the findings of this study are available from the corresponding author upon reasonable request. The scripts used to simulate the \gls{BBSS} suspension and for generating all figures in this paper are available online \cite{matlabscripts}.



\section*{References}
\label{sec:References} %

\bibliographystyle{aipauth4-1}


\printglossary[type=\acronymtype,style=long,nonumberlist]

\end{document}

%% file: glossary.tex
\newacronym{asc}{ASC}{Alignment Sensing and Control}
\newacronym{BBSS}{BBSS}{Big BeamSplitter Suspension}
\newacronym{BOSEM}{BOSEM}{Birmingham Optical Sensor and Electro-Magnetic actuator}
\newacronym{DARM}{DARM}{Differential ARM length}
\newacronym{dof}{dof}{degrees of freedom}
\newacronym{HOQI}{HoQI}{Homodyne Quadrature Interferometer}
\newacronym{ISI}{ISI}{Internal Seismic Isolation}
\newacronym{HEPI}{HEPI}{Hydraulic External Pre-Isolation}
\newacronym{L}{L}{longitudinal}
\newacronym{LHO}{LHO}{\acrshort{LIGO} Hanford Observatory}
\newacronym{LIGO}{LIGO}{Laser Interferometer Gravitational-Wave Observatory}
\newacronym{LLO}{LLO}{\acrshort{LIGO} Livingston Observatory}
\newacronym{LSC}{LSC}{Length Sensing and Control}
\newacronym{M1}{M1}{beamsplitter top mass}
\newacronym{M2}{M2}{beamsplitter intermediate mass}
\newacronym{M3}{M3}{beamsplitter optic}
\newacronym{P}{P}{pitch}
\newacronym{RMS}{RMS}{Root Mean Squared}
\newacronym{RX}{RX}{rotation around X axis}
\newacronym{RY}{RY}{rotation around Y axis}
\newacronym{SUSP}{SUSP}{SUSpension Point}
\newacronym{Y}{Y}{yaw}